\documentclass[aps,prl,twocolumn,groupedaddress,floatfix]{revtex4}

\usepackage[dvips]{graphicx,color} 
\usepackage{array,hhline,dcolumn} 
\usepackage{rotating} 
\usepackage{verbatim}

\newcommand{\chisq}{\chi^2}
\newcommand{\dchisq}{\Delta \chisq}

\newcommand{\dmsq}{\Delta m^{2}}

\newcommand{\numu}{\nu_{\mu}}
\newcommand{\numubar}{\bar{\nu}_{\mu}}
\newcommand{\nue}{\nu_e}
\newcommand{\nuebar}{\bar{\nu}_e}

\newcommand{\sinsqtheta}{\sin^2 2 \theta}

\newcommand{\evsq}{\mathrm{{eV}^{2}}}

\newcommand{\gtwid}{\mathrel{\raise.3ex\hbox{$>$\kern-.75em\lower1ex\hbox{$\sim$}}}}
\newcommand{\ltwid}{\mathrel{\raise.3ex\hbox{$<$\kern-.75em\lower1ex\hbox{$\sim$}}}}

\begin{document}

\title{Compatibility of high-$\Delta \mathrm{m}^2$ $\nue$ and $\nuebar$ neutrino oscillation searches}

\author{A.~A. Aguilar-Arevalo$^{5}$, C.~E.~Anderson$^{16}$, A.~O.~Bazarko$^{12}$,
        S.~J.~Brice$^{7}$, B.~C.~Brown$^{7}$,
        L.~Bugel$^{5}$, J.~Cao$^{11}$,
        L.~Coney$^{5}$,
        J.~M.~Conrad$^{5}$, D.~C.~Cox$^{8}$, A.~Curioni$^{16}$,
        Z.~Djurcic$^{5}$, D.~A.~Finley$^{7}$, B.~T.~Fleming$^{16}$,
        R.~Ford$^{7}$, F.~G.~Garcia$^{7}$,
        G.~T.~Garvey$^{9}$, C.~Green$^{7,9}$, J.~A.~Green$^{8,9}$, 
        T.~L.~Hart$^{4}$, E.~Hawker$^{3,9}$,
        R.~Imlay$^{10}$, R.~A. ~Johnson$^{3}$, G.~Karagiorgi$^{5}$,
        P.~Kasper$^{7}$, T.~Katori$^{8}$, T.~Kobilarcik$^{7}$,
        I.~Kourbanis$^{7}$, S.~Koutsoliotas$^{2}$, E.~M.~Laird$^{12}$, S.~K.~Linden$^{16}$,
        J.~M.~Link$^{14}$, Y.~Liu$^{11}$,
        Y.~Liu$^{1}$, W.~C.~Louis$^{9}$,
        K.~B.~M.~Mahn$^{5}$, W.~Marsh$^{7}$, P.~S.~Martin$^{7}$,
G.~McGregor$^{9}$,
        W.~Metcalf$^{10}$, P.~D.~Meyers$^{12}$,
        F.~Mills$^{7}$, G.~B.~Mills$^{9}$,
        J.~Monroe$^{5}$, C.~D.~Moore$^{7}$, R.~H.~Nelson$^{4}$, V.~T.~Nguyen$^{5}$,
        P.~Nienaber$^{13}$, S.~Ouedraogo$^{10}$, R.~B.~Patterson$^{12}$,
        D.~Perevalov$^{1}$, C.~C.~Polly$^{8}$, E.~Prebys$^{7}$,
        J.~L.~Raaf$^{3}$, H.~Ray$^{9,15}$, B.~P.~Roe$^{11}$,
A.~D.~Russell$^{7}$,
        V.~Sandberg$^{9}$, R.~Schirato$^{9}$,
        D.~Schmitz$^{5}$, M.~H.~Shaevitz$^{5}$, F.~C.~Shoemaker$^{12}$,
        D.~Smith$^{6}$, M.~Soderberg$^{16}$,
        M.~Sorel$^{5}$,
        P.~Spentzouris$^{7}$, I.~Stancu$^{1}$,
        R.~J.~Stefanski$^{7}$, M.~Sung$^{10}$, H.~A.~Tanaka$^{12}$,
        R.~Tayloe$^{8}$, M.~Tzanov$^{4}$,
        R.~Van~de~Water$^{9}$, M.~O.~Wascko$^{10}$, D.~H.~White$^{9}$,
        M.~J.~Wilking$^{4}$, H.~J.~Yang$^{11}$,
        G.~P.~Zeller$^{5,9}$, E.~D.~Zimmerman$^{4}$ \\
\smallskip
(The MiniBooNE Collaboration) 
\smallskip
}
\smallskip
\smallskip
\affiliation{
$^1$University of Alabama; Tuscaloosa, AL 35487 \\
$^2$Bucknell University; Lewisburg, PA 17837 \\
$^3$University of Cincinnati; Cincinnati, OH 45221\\
$^4$University of Colorado; Boulder, CO 80309 \\
$^5$Columbia University; New York, NY 10027 \\
$^6$Embry Riddle Aeronautical University; Prescott, AZ 86301 \\
$^7$Fermi National Accelerator Laboratory; Batavia, IL 60510 \\
$^8$Indiana University; Bloomington, IN 47405 \\
$^9$Los Alamos National Laboratory; Los Alamos, NM 87545 \\
$^{10}$Louisiana State University; Baton Rouge, LA 70803 \\
$^{15}$University of Florida; Gainesville, FL 32611 \\
$^{11}$University of Michigan; Ann Arbor, MI 48109 \\
$^{12}$Princeton University; Princeton, NJ 08544 \\
$^{13}$Saint Mary's University of Minnesota; Winona, MN 55987 \\
$^{14}$Virginia Polytechnic Institute \& State University; Blacksburg, VA 24061\\
$^{16}$Yale University; New Haven, CT 06520\\
}

\date{\today}

\begin{abstract}

This article presents the compatibility of experimental 
data from neutrino oscillation experiments with a high-$\dmsq$ two-neutrino oscillation 
hypothesis.  Data is provided by the Bugey, Karlsruhe 
Rutherford Medium Energy Neutrino Experiment 2 (KARMEN2), 
Los Alamos Liquid Scintillator Neutrino Detector (LSND), and MiniBooNE experiments.  
The LSND, KARMEN2, and MiniBooNE results are 25.36\% compatible within a two-neutrino 
oscillation hypothesis.  However, the point of maximal compatibility is found in a region 
that is excluded by the Bugey data.  
A joint analysis of all four experiments, performed 
in the $\sinsqtheta\ \mathrm{vs} \ \dmsq$ region common to all data,
 finds a maximal compatibility of 3.94\%.  This result does not account for additions to 
the neutrino oscillation model from sources such as CP violation or sterile neutrinos.
\end{abstract}

\pacs{14.60.Lm, 14.60.Pq, 14.60.St}

\keywords{Suggested keywords}
\maketitle

\section{Introduction}

Neutrino oscillations have been reported at three different $\dmsq$ scales: solar, 
atmospheric, and high-$\Delta m^2$.  
The solar~\cite{solar} and atmospheric~\cite{atm} best-fit results have been observed by several 
independent experiments, using various neutrino sources and techniques.  The high-$\Delta m^2$
 result, from the LSND detector~\cite{lsnd}, has yet to be reproduced.  

The LSND experiment observed a significant excess of events which 
are best fit by $\numubar \rightarrow \nuebar$ oscillations 
at the $\dmsq \sim 1$ $\evsq$ scale. The solar and atmospheric best fit results are at
$\dmsq \sim 8^{+0.6}_{-0.4}\times 10^{-5}$ $\evsq$ and $\dmsq \sim 2.4^{+0.6}_{-0.5}\times 10^{-3}$ $\evsq$, respectively.  
This wide spread of $\dmsq$ scales cannot be accommodated by the three neutrino mass states of the 
standard model.  The LSND result is uniquely incompatible with other 
oscillation observations, and if verified would demand extensions to the standard 
model in the neutrino sector ~\cite{mich}~\cite{georgia}.  

Prior to 2007, two experiments, KARMEN2 and Bugey, performed 
searches for oscillations in the region of oscillation parameter space probed by LSND.  
KARMEN2~\cite{karmen} conducted an accelerator-based $\numubar \rightarrow \nuebar$ appearance search.
The Bugey~\cite{bugey} reactor experiment probed for oscillations using $\nuebar$
disappearance.  Neither experiment found evidence for neutrino oscillations.  
However, a joint analysis \cite{joint} between LSND and KARMEN2 found both were compatible with 
a two-neutrino oscillation hypothesis at 64\% confidence level (CL), for $\dmsq > 0.2$ eV$^2$, 
in a region not covered by Bugey.

The MiniBooNE experiment, located at Fermi National Accelerator Laboratory, 
was designed to fully explore the LSND result.  
In 2007, MiniBooNE published results from a $\numu \rightarrow \nue$ appearance oscillation 
search~\cite{miniboone}.  MiniBooNE 
observed no significant excess of events in an energy range from 475 MeV to 3 GeV.  MiniBooNE is presently
 collecting anti-neutrino data, for use in an $\numubar \rightarrow \nuebar$ appearance oscillation 
search.  

This analysis presents results from the combination of LSND, MiniBooNE, KARMEN2, 
and Bugey.  It is motivated by a need to determine if the LSND excess may be the result of 
two-neutrino oscillations, in light of these three 
null experiments.  Results presented in this article make use of the MiniBooNE 
neutrino data set, and do not include the unpublished anti-neutrino data.  
The compatibility found in this analysis is valid within the framework of standard 2-neutrino 
oscillations.

\section{Input Data}

Data from each experiment are provided in a two-dimensional (2-D) grid of $\sinsqtheta\ \mathrm{vs} \ \dmsq$.
The value at each grid point represents the agreement between the observed data and a two-neutrino 
oscillation hypothesis, with a signal appropriate to the oscillation parameters at that point.  
The data sets come in several different formats (log likelihood (ln(L)), $\Delta$ln(L), $\chisq$), 
spanning different $\sinsqtheta\ \mathrm{vs} \ \dmsq$ ranges.  
An optimal compatibility calculation would make use of the absolute $\chisq$, as opposed to the 
$\Delta \chisq$, which is the change in $\chisq$ between each grid point and the experiment's best fit point.
However, we were unable to obtain the absolute $\chisq$ information from all input experiments.  
Therefore, our compatibility calculation can only make use of relative $\Delta \chisq$ information.
All input data is transformed into a $\Delta \chisq$ surface 
in $\sinsqtheta\ \mathrm{vs} \ \dmsq$ space, with common 
$\sinsqtheta\ \mathrm{vs} \ \dmsq$ binning.

The transformation from the input ln(L), $\Delta$ln(L) grid to a $\Delta \chisq$ grid is derived 
as follows.  In the technique of maximum likelihood fitting, a per-event probability $p(x_i|\alpha)$ is 
constructed, where $x_i$ are the event-measured quantities and $\alpha$ are the theoretical parameters. 
The goal is to maximize the likelihood, ${\cal L}(\alpha)$, the probability of all events in the sample, 
assuming the given model probability $p(x_i|\alpha)$ for each event, 

\begin{equation}
{\cal L}(\alpha) = \Pi_i \ p(x_i|\alpha).
\end{equation}

The technique of $\chi^2$ fitting is a special case of likelihood fitting 
in which the per event probability is a Gaussian distribution, 

\begin{equation}
p(x_i|\alpha) = \frac{1}{\sqrt{2 \pi} \sigma_i} \ e^{-\frac{(x_i-f(x_i|\alpha))^2}{2 \sigma_i^2} }.
\end{equation}

\noindent
Using this per-event probability in a likelihood fit, the log likelihood becomes 

\begin{equation}
- \ln {\cal L}(\alpha) = \sum_i \frac{( x_i - f(x_i|\alpha ))^2}{ 2 \sigma_i^2} + \sum_i \ln \sqrt{2\pi} \sigma_i.
\label{eqn:simply}
\end{equation}

\noindent
The second sum in the $\ln {\cal L}$ equation does not typically depend on the theory parameters $\alpha$.  
Equation~\ref{eqn:simply} is minimized by minimizing the familiar $\chi^2$ function, 

\begin{equation}
\chi^2(\alpha) = \sum_i \frac{(x_i - f(x_i | \alpha))^2 }{\sigma_i^2}.
\end{equation}

\vspace{1cm}
From the point of view of minimization, contours, and interpretation of results, 
there is an equivalence between the likelihood and the $\chi^2$ functions, given by 

\begin{equation}
-2 \ln {\cal L}(\alpha) = \chi^{2}(\alpha).
\label{eqn:equivalence}
\end{equation}

\noindent
To convert between the input ln(L) or $\Delta$ln(L) data and the $\chisq$ or $\Delta \chisq$ used in 
this analysis, the input data is multiplied by a factor of -2.  
The validity of this conversion technique has been verified by comparing calculated 
allowed regions found using the $\Delta$ln(L) and $\Delta \chisq$ grids, for LSND and KARMEN2,
 with those published by these experiments.

The input experiments published observation and limit curves using two different methods.  The first 
method is a two-dimensional (2D) global scan that calculates the $\Delta \chisq$ or $\Delta$ln(L) with respect to the global 
best fit point across the entire grid.  LSND and KARMEN2 calculated their results using this method.  The 
second method is a one-dimensional (1D) raster scan that calculates the change with 
respect to the local best fit point in each $\dmsq$ row.  The Bugey and MiniBooNE experiments used this 
method to produce their exclusion curves.  Given the mixture of methods used to report results from the input 
data, we have performed our compatibility calculation using both methods.  For each input experiment we 
create two $\Delta \chisq$ grids - one using the 2D global scan method, and one 
using the 1D raster scan method.  

The LSND data are provided as a 2D histogram of ln(L) values containing the decay-in-flight
 and the decay-at-rest results.  
The input grid covers 0.000313 to 1.01 in $\sin^2 2\theta$ and 0.0098 to 101.16 $\evsq$ in $\Delta m^2$.
The LSND ln(L) grid is first converted into a $\Delta$ln(L) grid, and then multiplied by -2 to 
produce a $\Delta \chi^2$ grid.
The conversions from ln(L) to $\Delta$ln(L) and $\Delta$ln(L) to $\Delta \chi^2$
are tested by calculating the 2D 
90\% and 99\% CL allowed regions by stepping away from the global best fit 
point a certain number of units in $\Delta$ln(L), $\Delta \chi^2$ space.  $\Delta$ln(L) units are 2.3 (90\% CL) and 4.6 (99\% CL); $\Delta \chi^2$ units are, for a 2D, 2 degree of freedom scan: 4.61 (90\% CL) and 9.21 (99\% CL).  Both tests properly reproduce the published LSND result~\cite{lsnd}.

The data from the KARMEN2 experiment are $\Delta$ln(L) values covering a range of
0.000316 to 1 ($\sin^2 2\theta$) and 0.01 to 100 $\evsq$ ($\Delta m^2$).
Each point is multiplied by $\mathrm{-2}$ to produce a $\Delta \chisq$ grid.  
A cross-check is performed using the $\Delta\chi^2$ grid, where 
the probability is calculated at each point on the grid 
using 2 degrees of freedom (DOF).  Points where the probability crosses 10\% delineate the 90\%  
exclusion band.  This test correctly finds the 90\% CL exclusion band from the KARMEN2 
publication~\cite{karmen}.

We were unable to obtain data directly from the Bugey collaboration.  
However, a recent global analysis of Gallium and reactor $\nu_e$ disappearance data 
describes a method used to reproduce the $\chi^2$ surface of Bugey, complete with full systematic errors~\cite{giunti}.
The authors kindly provided us with their full $\chi^2$ surface ($\sin^2 2\theta$: 0.01 to 1, 
$\Delta m^2$: 0.01 to 100 $\evsq$).  A cross-check was also performed on this data by applying the raster scan method 
and stepping away from each local best fit point by 2.71 $\Delta \chisq$ units to find the 90\% CL 
exclusion band.  Using this method we are able to reproduce the published Bugey result~\cite{bugey}.

The MiniBooNE data is expressed as a $\chi^2$ format from 0.0001 to 0.4108 in $\sin^2 2\theta$ and 0.0488 to 
51.13 $\evsq$ in $\Delta m^2$.  This analysis utilizes the MiniBooNE data from 475 MeV to 3 GeV in neutrino energy; 
the low energy region (below 475 MeV) is not considered in the compatibility calculation.

\section{Compatibility Calculation}

The compatibility calculation uses a method developed by Maltoni and Schwetz~\cite{msch} to answer the specific question, 
\emph{``How probable is it that all experimental results come from the same underlying 
two-neutrino oscillation hypothesis?''}.
First, a $\Delta \chisq$ grid is constructed for each experiment as described in the previous section.  The individual grids are then summed 
together to produce one summed $\Delta \chisq$ grid.
The compatibility test statistic, $\bar{\chi}^2_{min}$, 
is the minimum of the summed $\Delta \chisq$ grid.  
$\bar{\chi}^2_{min}$ follows a $\chi^2$ distribution 
with $P_c$ degrees of freedom, where $P_c$ is the sum of the total number of 
independent parameters minus the number of independent parameters estimated from the data.
For example, the combination of the 2D MiniBooNE and 2D LSND results yields four total independent 
parameters; each experiment independently measures $\sinsqtheta$ and $\dmsq$.  Two parameters 
are estimated from the data ($\sinsqtheta$,$\dmsq$), resulting in a $P_c$ of 2.
The final compatibility is the $\chisq$ probability of $\bar{\chi}^2_{min}$ using $P_c$ degrees of freedom.
In the analysis using two experiments $P_c$ is 2, for three experiments $P_c$ is 4, 
and for all four experiments $P_c$ is 6.

This method is designed to be robust against cases where the $\chi^2$ minima of the 
individual data sets are very low, and when several parameters are fitted to a large number of 
data points.  It reduces the problem that a possible disagreement between
data sets becomes diluted by data points which are insensitive to the crucial parameters.

Of course, there are limitations to this method.  This method does not take into account the 
absolute goodness of fit of each individual experiment at its own best fit point.
It is also valid only for truly statistically independent data sets.  Theoretical uncertainties in 
similar experiments may introduce correlations between the various results; for example, LSND and KARMEN2 have 
the same neutrino beam energy spectrum which may result in similar neutrino interaction errors.  
However, a previously reported combined analysis of LSND and KARMEN claims the 
two experiments may be considered independent~\cite{joint}.  MiniBooNE and Bugey are not expected to have any 
uncertainties in common with the other experiments.

\section{Allowed Region Calculation}

Combinations of experiments which result in a compatibility of greater than 10\% are further explored to locate any
remaining allowed regions.  Allowed regions are indicated by closed contours in the $\sin^2 2\theta\ \mathrm{vs} \ \Delta m^2$
plane.  Contours which do not close form exclusion bands; parameter values situated to the right of the bands are excluded at 
a given CL (typically 90 and 99\% CL).  The allowed regions indicate where the oscillation parameters would lie, at a given confidence level, 
assuming all experimental results can arise in a framework of two-neutrino oscillations.
The calculated compatibility is the metric for how valid this assumption is.

The allowed regions calculation follows the prescription of Roe~\cite{byron}.  Each experiment's
$\Delta \chisq$ grid is 
converted into a $\Delta \chisq$ probability grid, using an appropriate number of DOF (two 
for the global scan analysis, one for the raster scan analysis).
The final combined probability at a given point can be obtained from the 
product of the individual probabilities, \emph{x}.  The result is a sum of 
powers of the absolute value of the logarithms of \emph{x}, 

\begin{equation}
Prob = \emph{x} \cdot \sum^{n - 1}_{j = 0} \frac{1}{j!} \cdot |ln^j(\emph{x})|
\label{equation:Roe}
\end{equation}

\noindent
, where \emph{x} is the product of the individual probabilities and \emph{n} is the number of experiments being included.
Points where the probability crosses 10\% 
bound the 90\% confidence level allowed region; points where the probability crosses 
1\% bound the 99\% region.  

There has been much discussion regarding the number of DOF that one can use 
with the $\chi^2$ grids in the allowed region calculation, and how the DOF changes across the 
grids~\cite{ndfpaper}.  We examined the change in DOF across the 2D grid, using the Feldman-Cousins frequentist 
method~\cite{ndfpaper}.  Our study of the DOF finds that, in general, use of 2 DOF is valid across the 2D grid.  
This breaks down for points with high $\Delta m^2$ and high $\sin^2 2\theta$ values:
$\Delta m^2$ $>$ 10 $\evsq$ and $\sin^2 2\theta$ $>$ 0.01.  
However, the approximation of 2 DOF is still valid for $\sin^2 2\theta$ $>$ 0.01 and 
$\Delta m^2$ $<$ 10 $\evsq$.  The region where the 2 DOF approximation
is no longer valid is an area which does not contain any allowed regions from LSND, 
and as such should not impact this analysis.


The compatibility is defined as the $\dchisq$ probability at the best fit point.  The combination of 
experiments reduces the number 
of independent parameters used in the probability calculation.  The Maltoni-Schwetz method
can easily accommodate the change in degrees of freedom~\cite{msch}.  It is not clear how to include 
information about the number of degrees of freedom into the more traditional (Roe) method 
(Equation~\ref{equation:Roe}).  
The Roe method applied as-is results in too high a compatibility, due to the inability of the 
method to consider a reduced degree of freedom.  

We have chosen to use the more traditional Roe method to calculate the allowed regions, and the 
Maltoni-Schwetz method to find the compatibility.  Equivalently, we could have chosen to use the 
generic Maltoni-Schwetz method to find the allowed regions; both methods return identical values when 
evaluating the joint probability distribution function.

The range of $\sinsqtheta\ \mathrm{vs} \ \dmsq$ common to all experiments is used for the 
compatibility and allowed region calculations.  The $\dmsq$ is restricted to 0.0488 to 
51.13 $\evsq$ for all results.  The $\sinsqtheta$ range, for results without Bugey, is 0.000317 to 0.4108.  
Results containing Bugey employ a $\sinsqtheta$ range of 0.01 to 0.4108.

\section {How to Interpret the Results}

As previously discussed, the $\Delta \chisq$ grids used in this analysis are created in two ways.  
The 2D grids use the global best fit point internal to each experiment 
to produce the $\Delta \chisq$.  The compatibility extracted from the combined 2D $\Delta \chisq$ grid represents 
how probable it is that one could observe the experimental 
results if nature truly has two neutrino oscillations in this high-$\dmsq$ region (0.0488 to 
51.13 $\evsq$).  This method also
finds the most probable point for the true oscillations to exist across the evaluated phase space.  
The 1D results use a raster scan method to find the local best fit point at each $\dmsq$, 
internal to each experiment, to produce the $\Delta \chisq$.  The results from 
this method represent the compatibility at each $\dmsq$, if nature truly had 
two neutrino oscillations located at that $\dmsq$.

\section{Results}

The 2D analysis reports a single value for the maximum compatibility of the experimental data 
with the two neutrino oscillation hypothesis.  These results are presented in Table~\ref{table:maxcompat_2d}.  
The compatibility for the 1D analysis is a function of $\Delta m^2$, and is presented in graphical form in the 
following sections.  
All results are calculated with respect to an oscillation hypothesis valid
 in the $\dmsq$ region of 0.0488 to 51.13 $\evsq$.  This region has been further divided into three components 
; low $\dmsq$ indicates the region from 0.0488 to $\sim$1 $\evsq$, medium $\dmsq$ spans 
$\sim$1 to $\sim$7 $\evsq$, and high $\dmsq$ is $>\sim$7 $\evsq$.  These divisions are used to 
characterize the results in the following sections.

We first present results from the combination of the three accelerator-based appearance 
oscillation experiments (LSND, KARMEN2, MiniBooNE).  This combination finds a high compatibility (25.36\%) at
low $\dmsq$ (Figure~\ref{fig:mbkarlsndregions_2d}).  The inclusion of the Bugey reactor disappearance data highly constrains the low $\dmsq$ region, 
reducing the compatibility to a low level (3.94\%, Figure~\ref{fig:mbkarlsndbugprob_2d_zoom}).  
In the second section, 
the compatibility and allowed regions are explored in 
various combinations of the null experiments (KARMEN2, MiniBooNE, and Bugey).  In the third section, 
the result of 2.14\% compatibility omits the KARMEN2 data.  The final section discusses the combination of LSND 
and KARMEN2.

\begin{table}[ht]
\footnotesize
\begin{center}
\begin{tabular}{|c|c|c|c|c|c|c|} \hline
 \multicolumn{1}{c}{LSND} &  \multicolumn{1}{c}{KARMEN2} &  \multicolumn{1}{c}{MB}
& \multicolumn{1}{c}{Bugey} & \multicolumn{1}{c}{Max Compat (\%)} 
& \multicolumn{1}{c}{$\Delta m^2$} & \multicolumn{1}{c}{$\sin^2 2\theta$} \\ \hline \hline

X & X & X &    & 25.36& 0.072 & 0.256 \\ \hline
X & X & X &  X &  3.94 & 0.242 & 0.023 \\ \hline

\multicolumn{7}{|c|}{} \\ \hline 

 & X & X &    & 73.44 & 0.052 & 0.147 \\ \hline
 & X & X &  X & 27.37 & 0.221 & 0.012 \\ \hline

\multicolumn{7}{|c|}{} \\ \hline 

X & & X &    & 16.00 & 0.072 & 0.256 \\ \hline
X & & X &  X &  2.14 & 0.253 & 0.023 \\ \hline

\multicolumn{7}{|c|}{} \\ \hline 

X & X & &    & 32.21 & 0.066 & 0.4 \\ \hline

\hline 

\end{tabular}
\normalsize
\caption{Maximum compatibility for a variety of combinations of the input experiments, found 
using the 2D $\Delta \chi^2$ grids.  The last two columns indicate the $\dmsq\ \mathrm{vs} \ \sinsqtheta$ location of the 
point of maximum compatibility.  The X's indicate which experiments were included in the analysis.}
\label{table:maxcompat_2d}
\end{center}
\end{table}

\subsection{LSND, KARMEN2, MiniBooNE}

\begin{figure}[h]
\scalebox{0.35}{\includegraphics[angle=0]{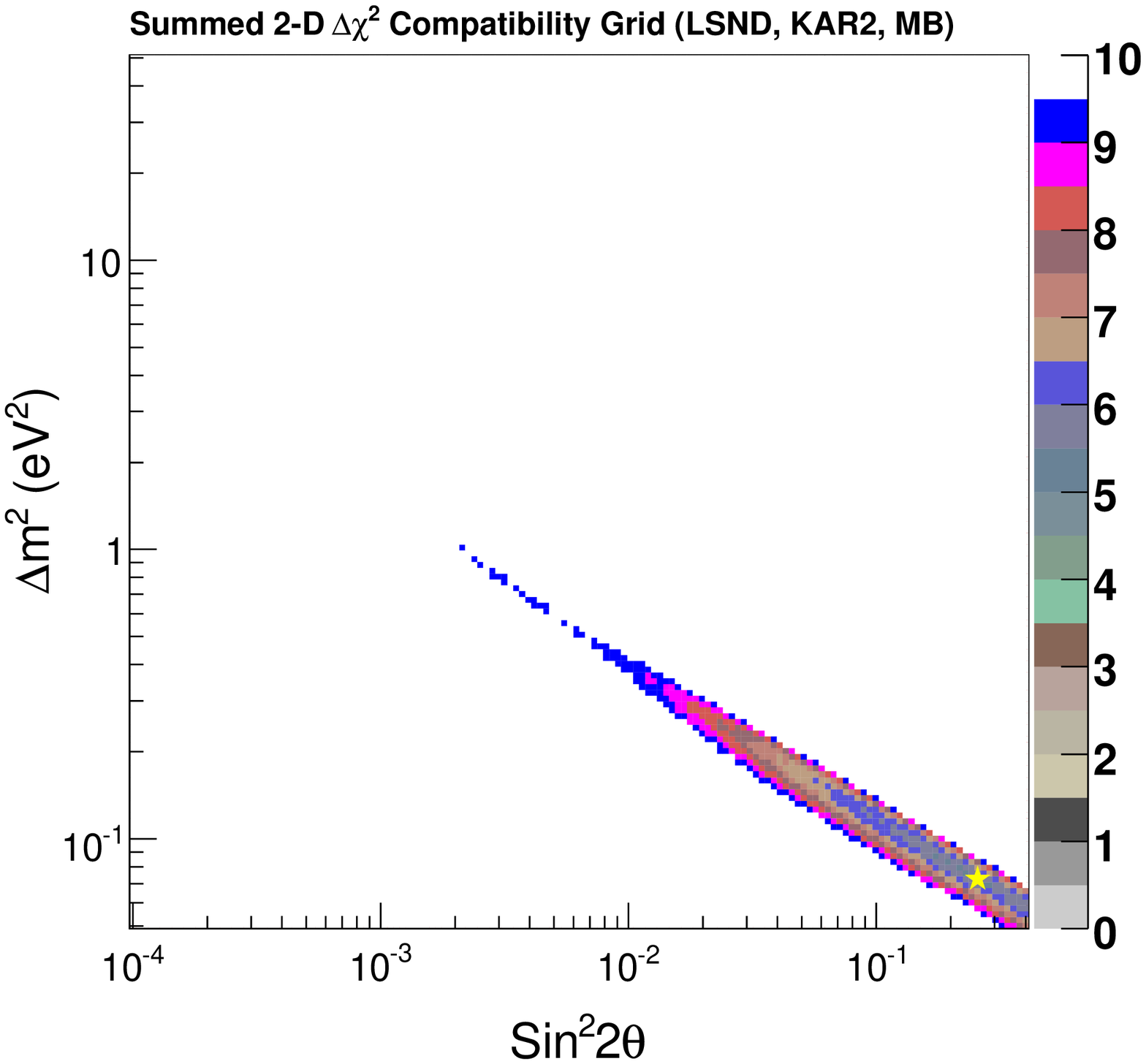}}
\hfill
\scalebox{0.35}{\includegraphics[angle=0]{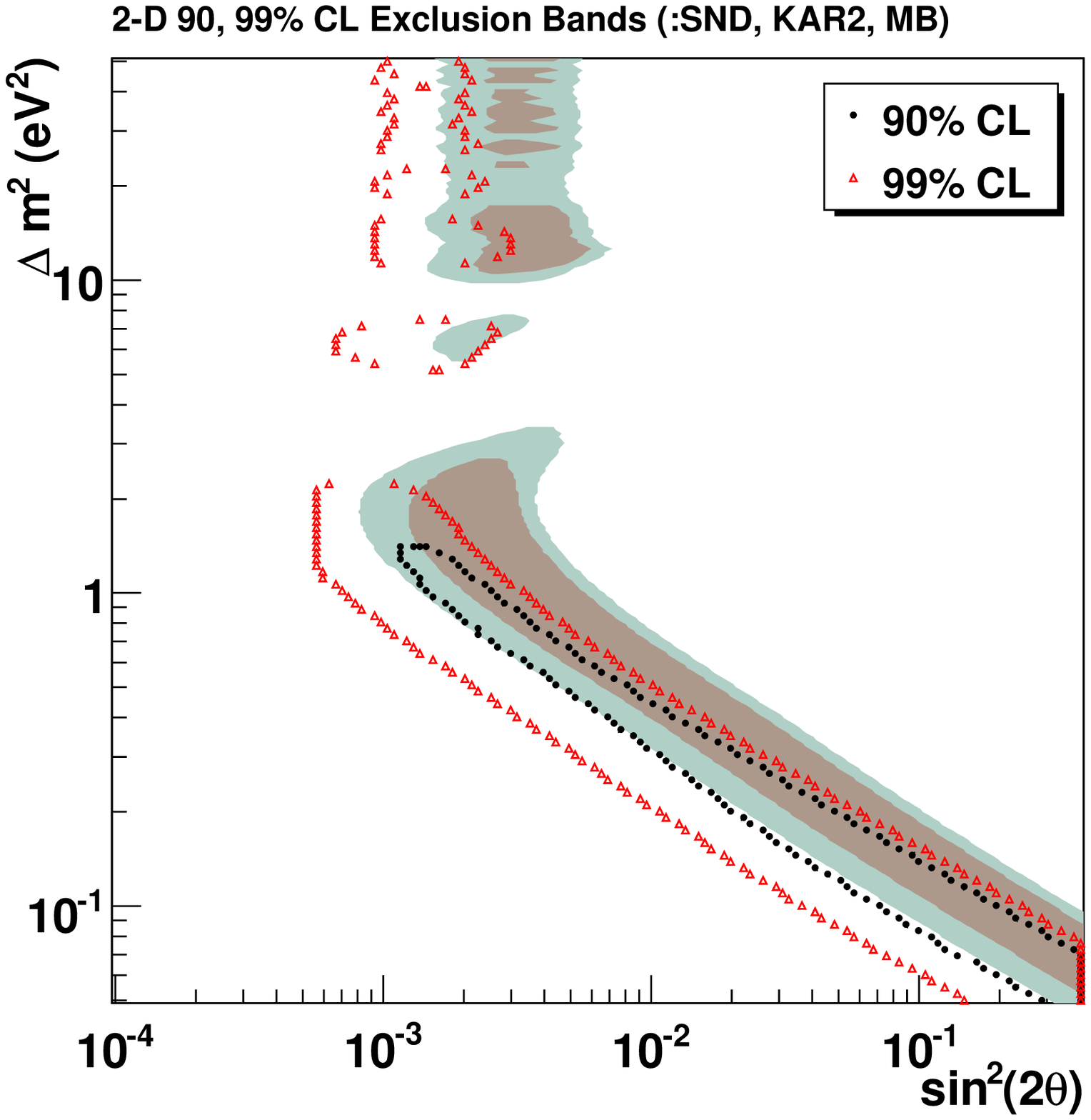}}
\caption{Top: Summed 2D $\Delta\chi^2$ compatibility grid from LSND, KARMEN2, and MiniBooNE.  The star indicates the point of maximal compatibility (25.36\%).  Bottom: Allowed regions (90\%, 99\%) found for the 2D LSND, KARMEN2, and MiniBooNE joint analysis.  Triangle points contain the 99\% CL region, circle points contain the 90\% CL region.  The solid brown area is the LSND 90\% allowed region; the solid light blue area is the LSND 99\% allowed region.  The vertical straight edge on the left arises from a sharp discontinuity in the LSND input grid.}

\label{fig:mbkarlsndregions_2d}

\end{figure}

First, we consider only results from the $\nue$ appearance searches.  
Figure~\ref{fig:mbkarlsndregions_2d} (top) displays the 2D $\Delta \chisq$ grid from the combination of 
 LSND, KARMEN2, and MiniBooNE.  The point of maximal compatibility (25.36\%) is indicated by the star.  
The point of highest compatibility is not limited by the $\sinsqtheta\ \mathrm{vs} \ \dmsq$ grid boundaries, but is 
found in a region excluded by the Bugey data.  The allowed regions for two-neutrino oscillations are 
shown in the bottom of Figure~\ref{fig:mbkarlsndregions_2d}.  
There are 99\% allowed regions at low, medium, and 
high $\dmsq$.  The only 90\% allowed region is located at low $\dmsq$, which overlaps slightly with the LSND 
90\% allowed region. 

\begin{figure}[h]
\scalebox{0.35}{\includegraphics[angle=0]{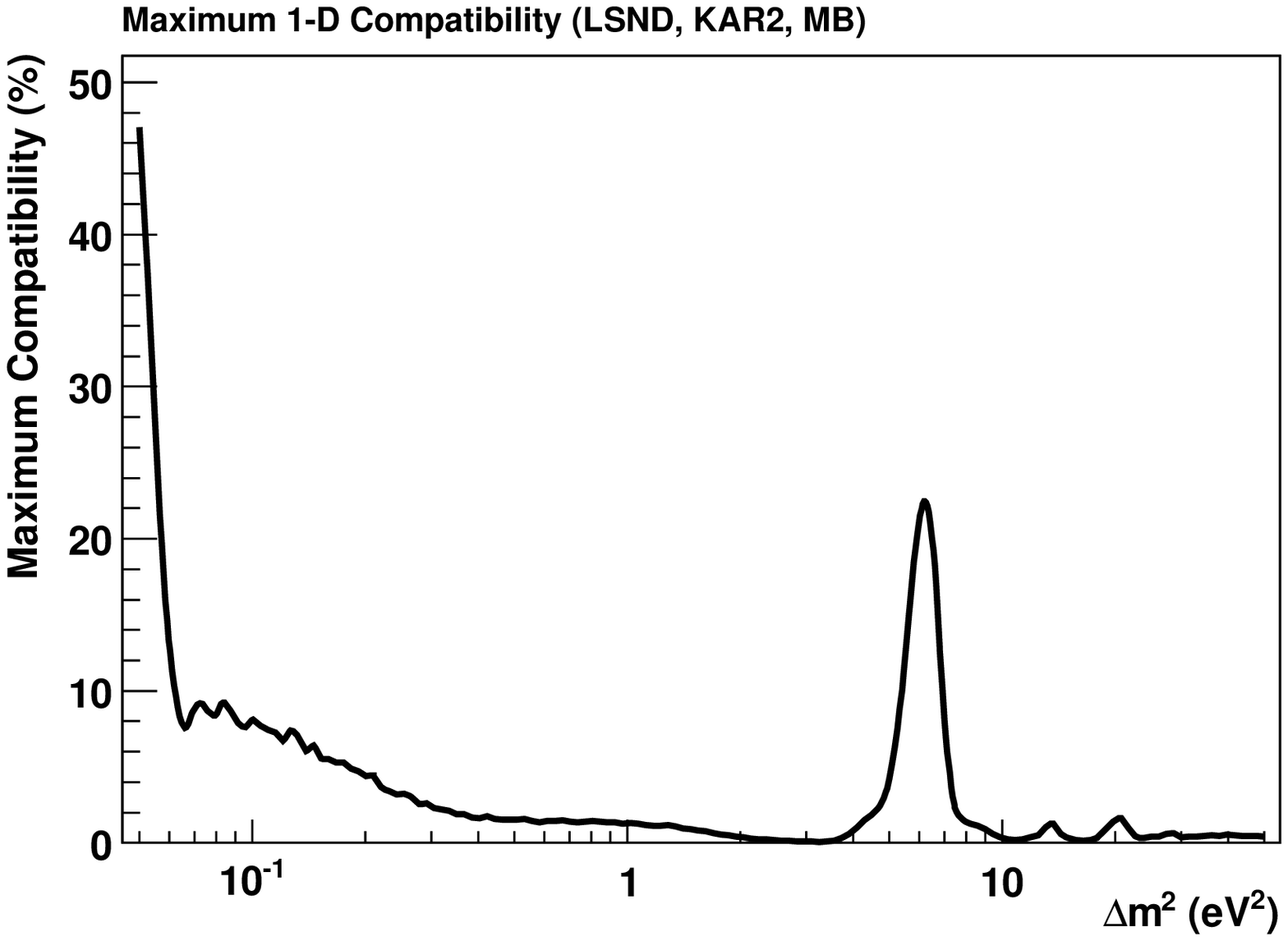}}
\hfill
\scalebox{0.35}{\includegraphics[angle=0]{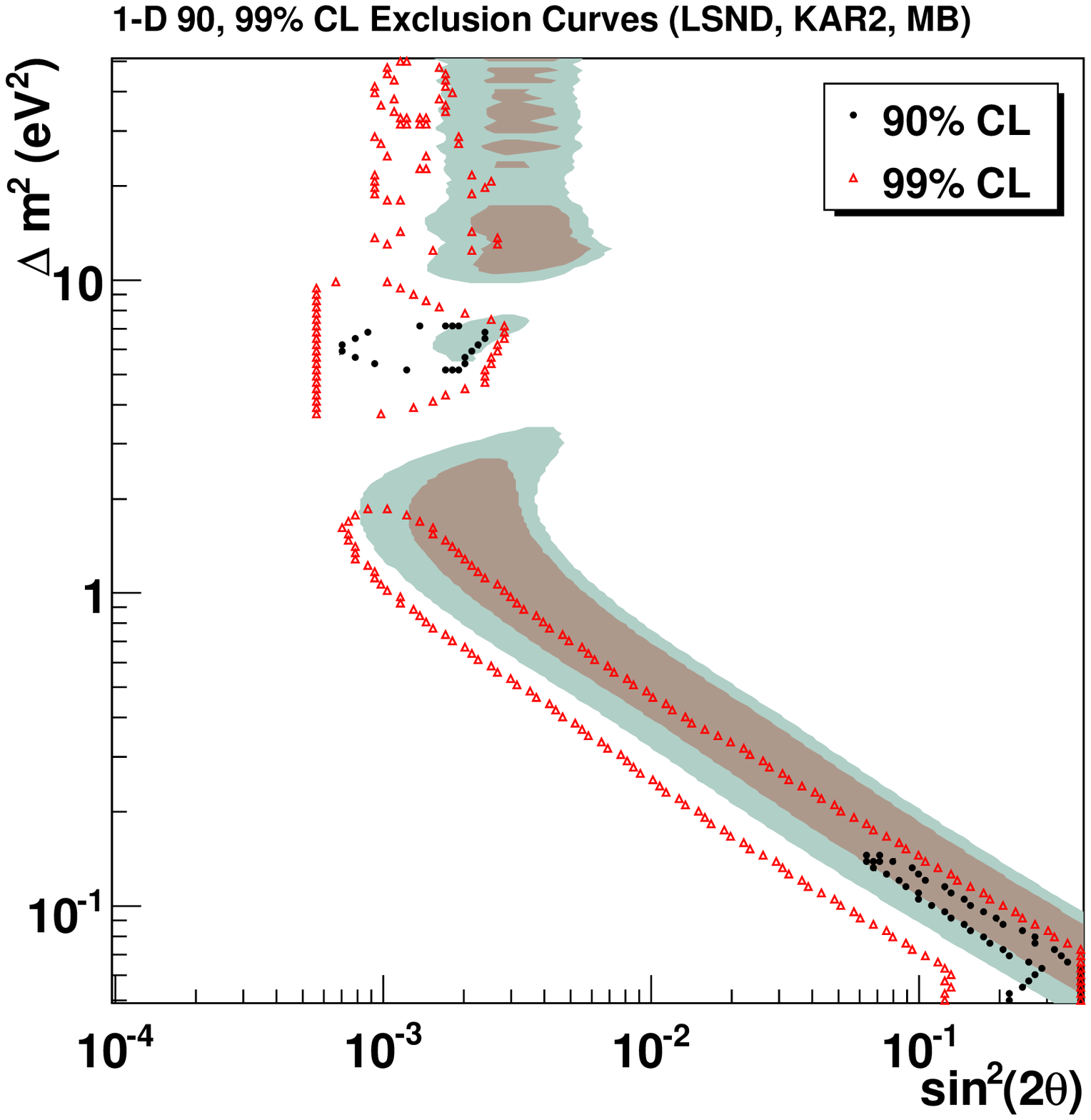}}
\caption{Top: Maximum compatibility as a function of $\Delta m^2$ for the 1D LSND, KARMEN2, MiniBooNE analysis.  Bottom: Allowed regions for the 1D LSND, KARMEN2, MiniBooNE analysis.  90\% allowed regions exist in the low and mid $\dmsq$ regions.  Triangle points contain the 99\% CL region, circle points contain the 90\% CL region.  The vertical straight edge on the left arises from a sharp discontinuity in the LSND input grid.}
\label{fig:mblsndkarregions_1d}
\end{figure}

Compatibility values are also reported for the 1D analysis as a function of $\dmsq$.  
The maximum compatibility for the 1D 
LSND, KARMEN2, MiniBooNE analysis is shown in the top of Figure~\ref{fig:mblsndkarregions_1d}.
The 1D analysis also finds a high compatibility (here almost 50\%) at low $\dmsq$.  In addition, 
this raster scan method allows for a $\sim$25\% compatible region at medium $\dmsq$.  
Figure~\ref{fig:mblsndkarregions_1d} (bottom)
 illustrates the 90 and 99\% allowed regions.  While these regions 
appear to be shifted from the LSND signal region, it must be remembered that the LSND signal region shown 
is that found using a 2D analysis, \emph{not} a 1D scan.  

\begin{figure}[h]
\scalebox{0.35}{\includegraphics[angle=0]{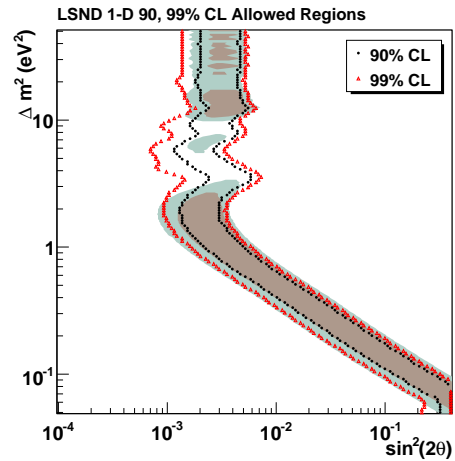}}
\caption{90 and 99\% allowed regions for the 1D scan of LSND, overlayed on the 2D published allowed regions.  Triangle points contain the 99\% CL region, circle points contain the 90\% CL region.}
\label{fig:lsndregions_1d}
\end{figure}

Finally, Figure~\ref{fig:lsndregions_1d} compares the LSND allowed 
regions found using the 1D $\Delta\chi^2$ method to those published by the LSND collaboration, found using 
the 2D scan.  
The 1D allowed regions are quite large in comparison to the 2D regions, and include areas which had 
no 90\% allowed islands in the 2D scan.

\subsection{LSND, KARMEN2, MiniBooNE, Bugey}

\begin{figure}[h]
\scalebox{0.35}{\includegraphics[angle=0]{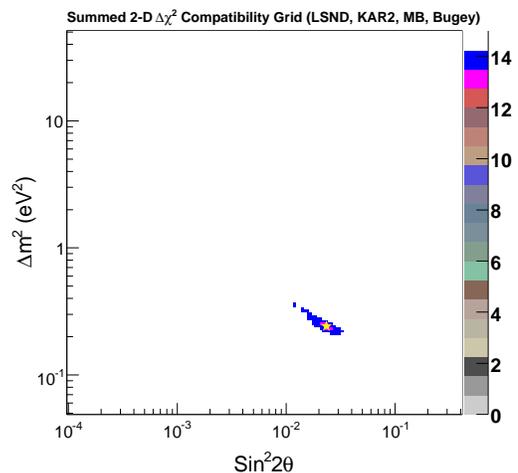}}
\caption{Summed 2D $\Delta\chi^2$ compatibility grid from LSND, KARMEN2, MiniBooNE, Bugey.  The star indicates the point of maximal compatibility (3.94\%).}
\label{fig:mbkarlsndbugprob_2d_zoom}
\end{figure}

The inclusion of the Bugey data significantly changes the compatibility results.  The Bugey $\sinsqtheta$ range 
has a lower bound at 0.01; analyses including Bugey are restricted to $\sinsqtheta$ of 
0.01 to 0.4108.  The combination of all four experiments has a 2D compatibility of 3.94\%.  
Figure~\ref{fig:mbkarlsndbugprob_2d_zoom} shows the final 2D $\Delta \chisq$ grid, for all four experiments.

\begin{figure}[h]
\scalebox{0.35}{\includegraphics[angle=0]{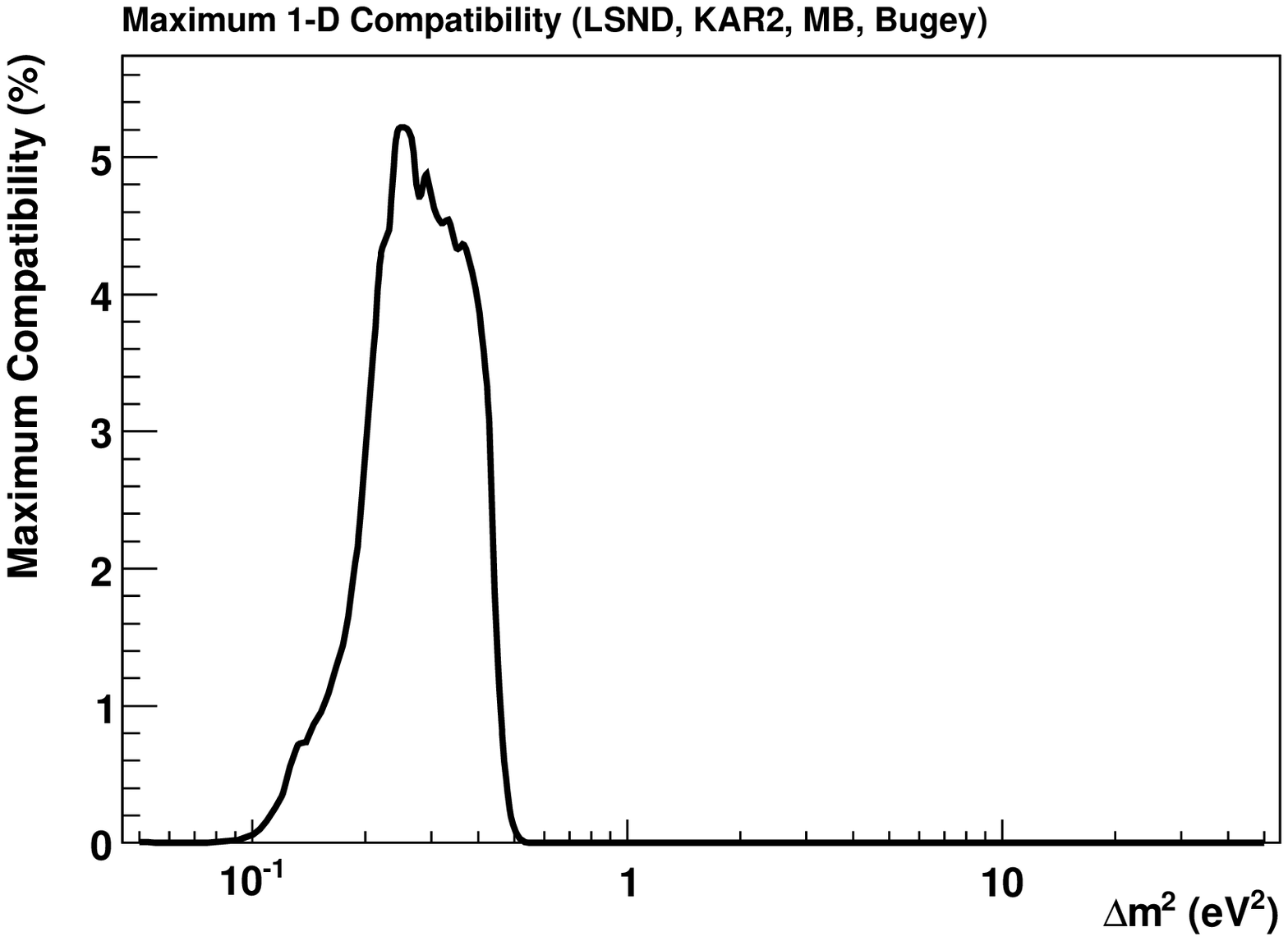}}
\caption{Maximum compatibility as a function of $\Delta m^2$ for the 1D LSND, KARMEN2, MiniBooNE, Bugey analysis.}
\label{fig:mblsndkarbugcomapt_1d}
\end{figure}

The 1D analysis of all four experiments agrees quite well with the 2D results; 
the point of highest compatibility is found in the low $\dmsq$ region, and all results are no more than 5.2\% 
compatible with having resulted from two-neutrino oscillations (Figure~\ref{fig:mblsndkarbugcomapt_1d}).

\subsection{KARMEN2, MiniBooNE}

\begin{figure}[h]
\scalebox{0.35}{\includegraphics[angle=0]{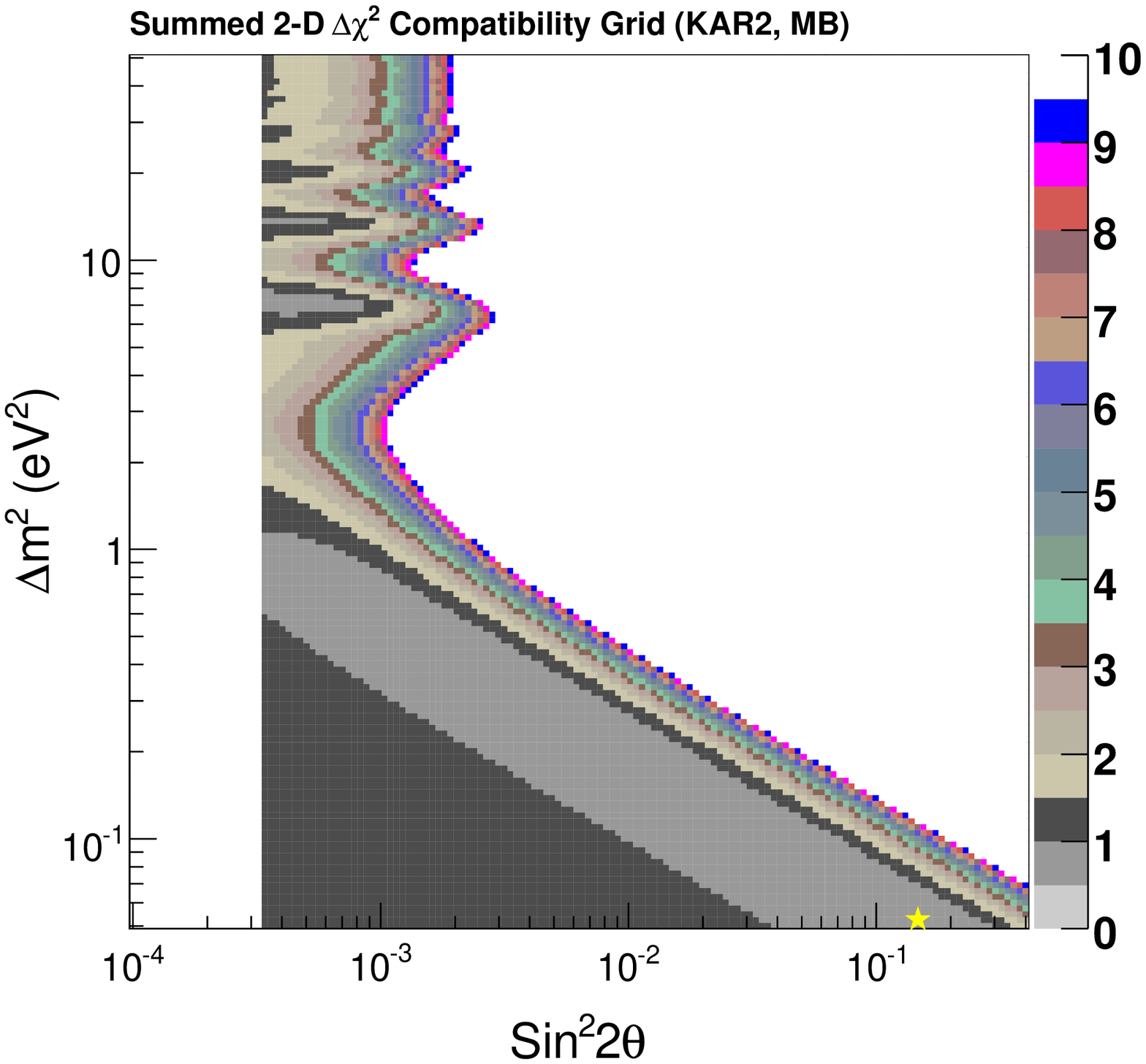}}
\hfill
\scalebox{0.35}{\includegraphics[angle=0]{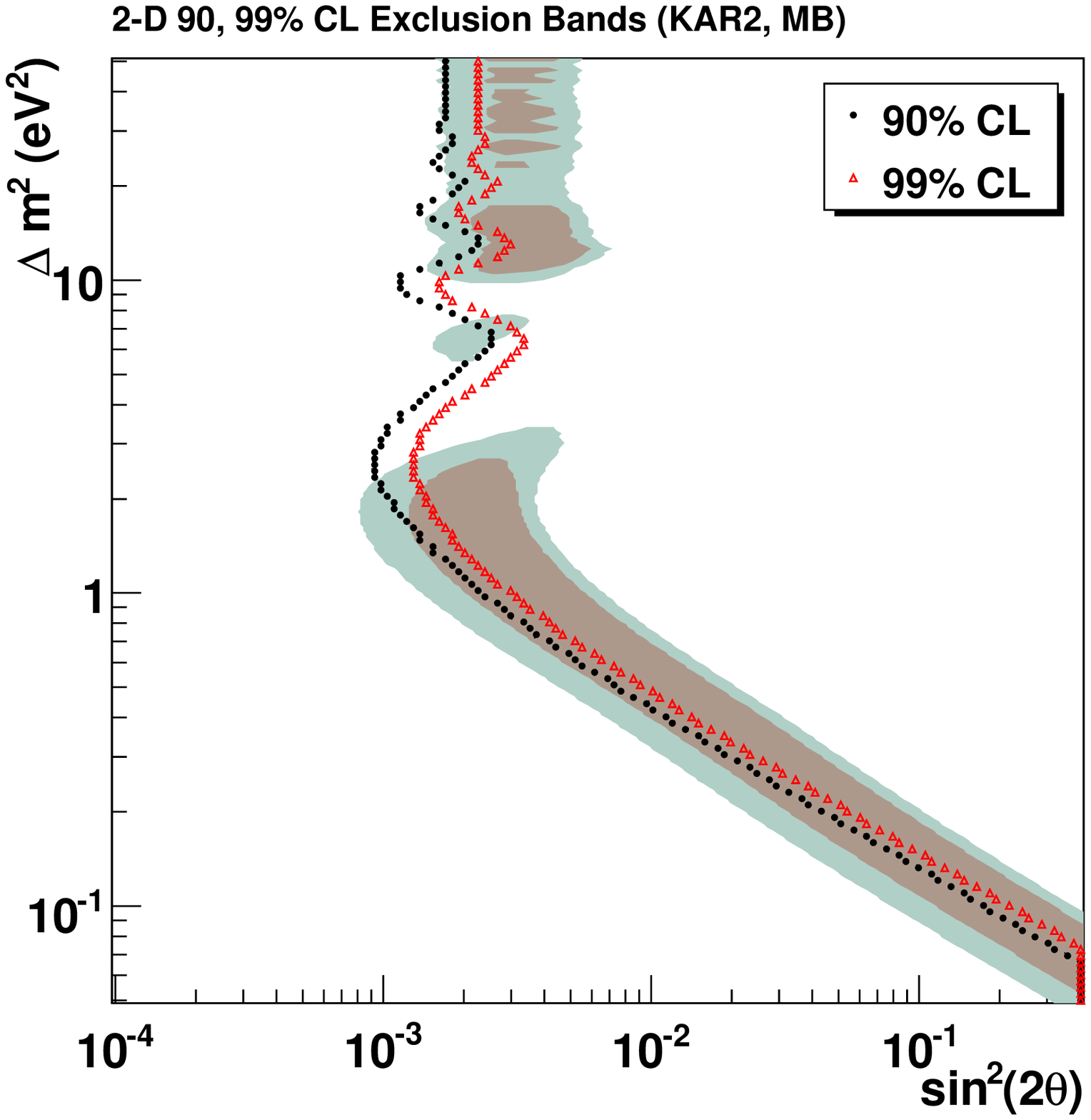}}
\caption{Top: Summed $\Delta\chi^2$ compatibility of KARMEN2 and MiniBooNE using the 2D analysis.  
The star indicates the point of maximal compatibility (73.44\%).  
The compatibility is limited by the boundaries of the analysis and may increase with a 
loosening of the grid range.  Bottom: Exclusion bands (90\%, 99\%) found for the 2D 
KARMEN2 and MiniBooNE joint analysis.  Values to the right of the lines are excluded 
at the 90, 99\% CL.  Triangle points form the 99\% CL band, circle points form the 90\% CL band.}
\label{fig:mbkarregions_2d}
\end{figure}

It is instructive to calculate compatibility and remaining allowed regions, in the absence of a positive 
LSND signal.  The 2D analysis finds KARMEN2 and MiniBooNE are 73.44\% compatible with a 
two-neutrino hypothesis; there is a 73.44\% chance that we would find these two null results, in the 
presence of two-neutrino oscillations at these $\sinsqtheta$, $\dmsq$ values (0.147, 0.052 $\evsq$).  
However, the minimum in the $\dchisq$ is outside of the region where either experiment has much sensitivity.  For example, the compatibility at the lowest grid point 
(0.0003, 0.05 $\evsq$) still 
remains high at 53.52\%.
The summed $\dchisq$ and allowed regions for oscillations 
are shown in Figure~\ref{fig:mbkarregions_2d}.

\begin{figure}[h]
\scalebox{0.35}{\includegraphics[angle=0]{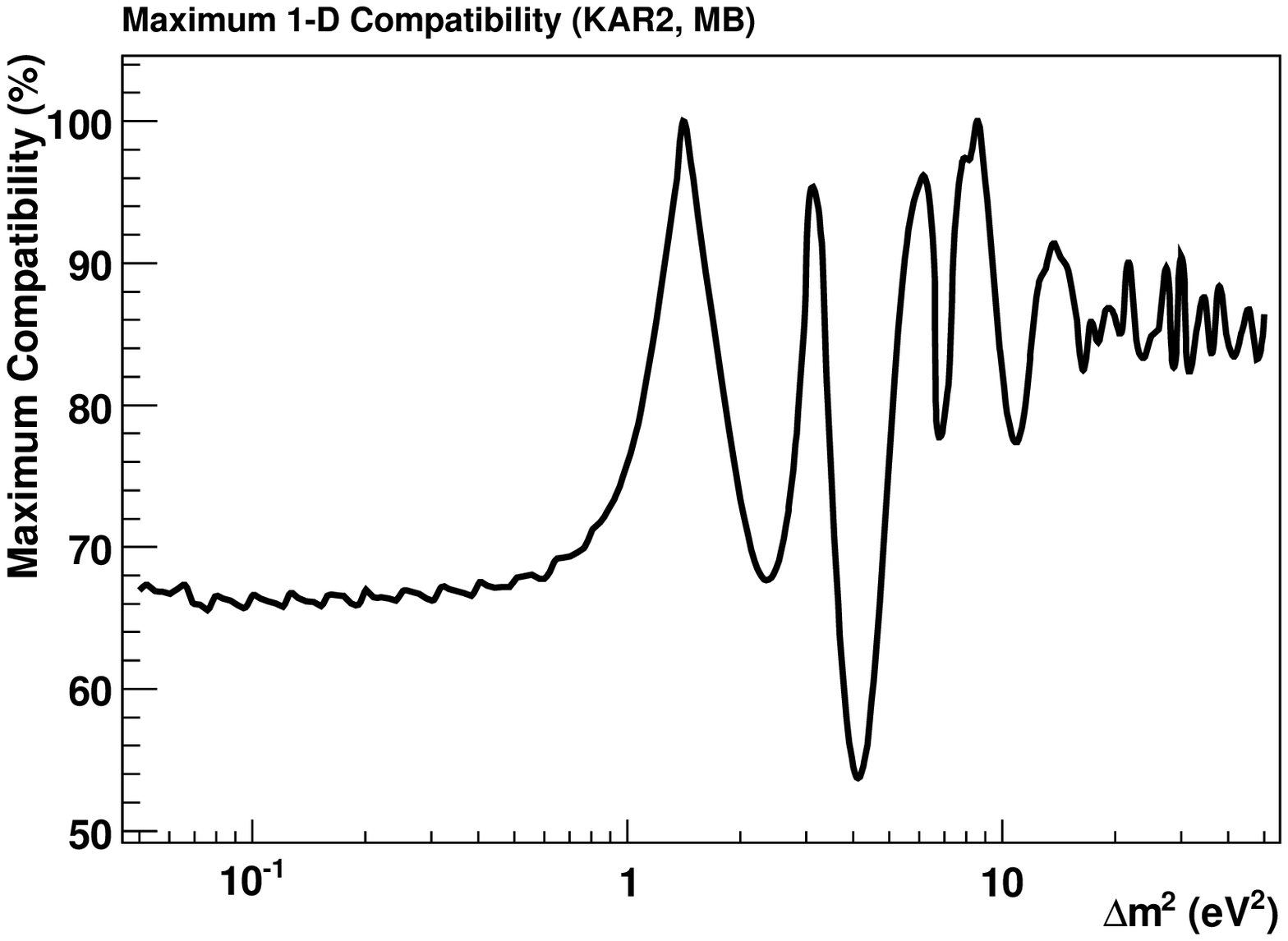}}
\hfill
\scalebox{0.35}{\includegraphics[angle=0]{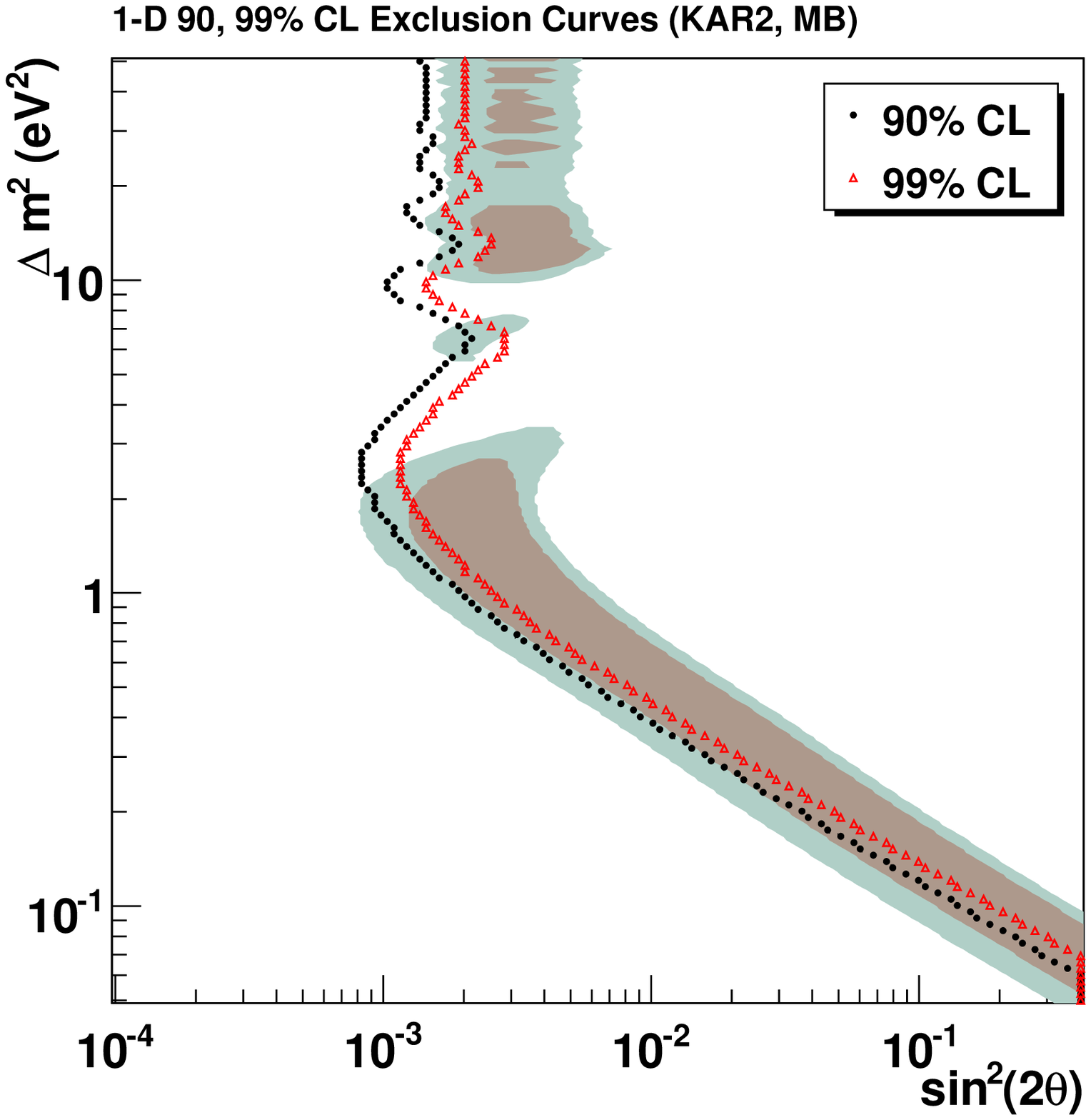}}
\caption{Top: Maximum compatibility as a function of $\Delta m^2$ for the 1D KARMEN2 and MiniBooNE analysis.  Bottom: Exclusion bands (90\%, 99\%) found for the 1D KARMEN2, MiniBooNE joint analysis.  Values to the right of the lines are excluded at the 90, 99\% CL.  Triangle points form the 99\% CL band, circle points form the 90\% CL band.  These curves are very similar to those found using the 2D method in Figure~\ref{fig:mbkarregions_2d}.}
\label{fig:mbkarregions_1d}
\end{figure}

The 1D analysis of KARMEN2 and MiniBooNE finds high compatibility that reaches 100\% at 
low and medium $\dmsq$ (top, Figure~\ref{fig:mbkarregions_1d}).  However, the 90 and 99\% CL exclusion curves 
(bottom, Figure~\ref{fig:mbkarregions_1d}) are almost identical to those found in the 2D analysis.

\subsection{KARMEN2, MiniBooNE, Bugey}

\begin{figure}[h]
\scalebox{0.35}{\includegraphics[angle=0]{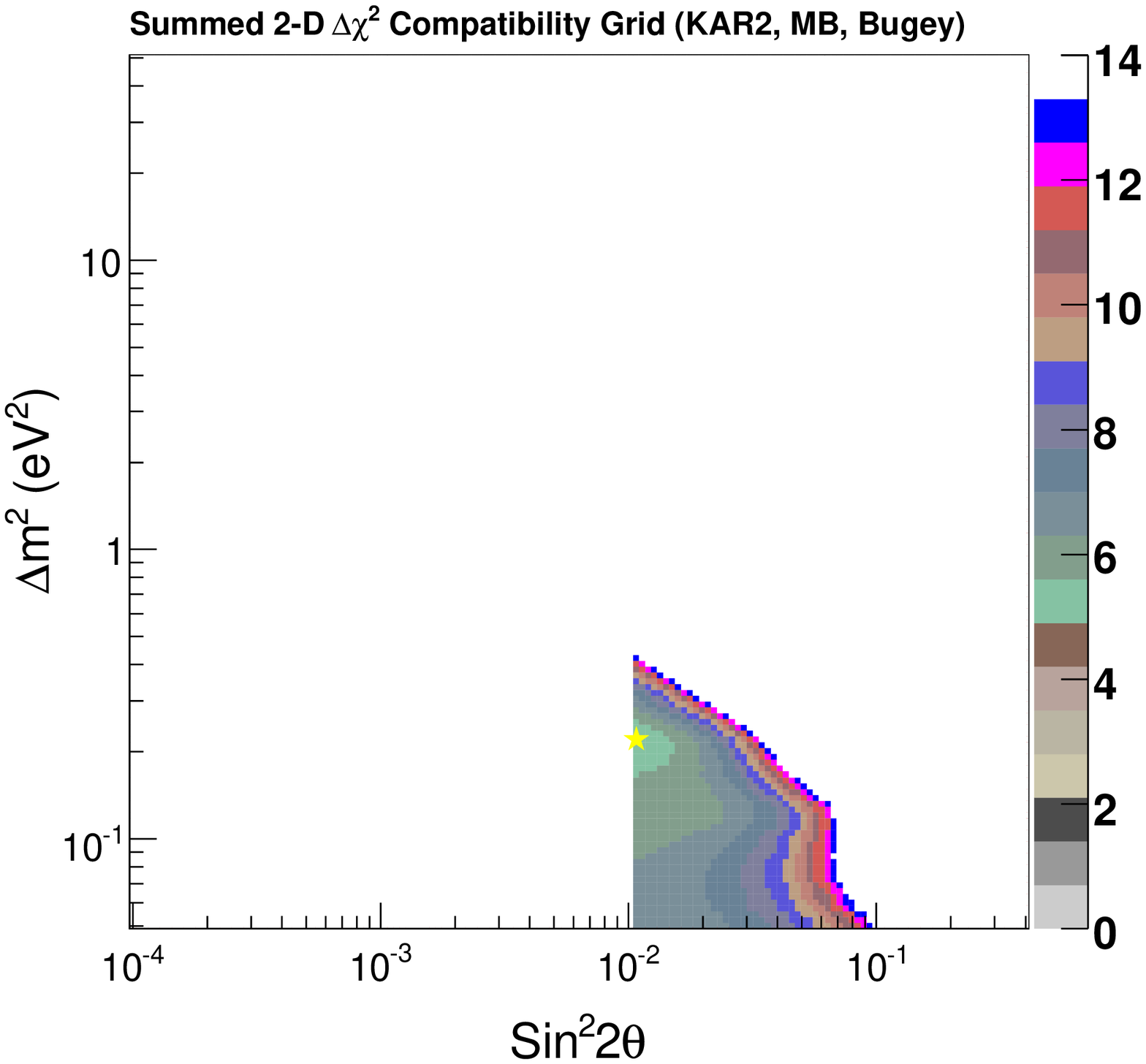}}
\hfill
\scalebox{0.35}{\includegraphics[angle=0]{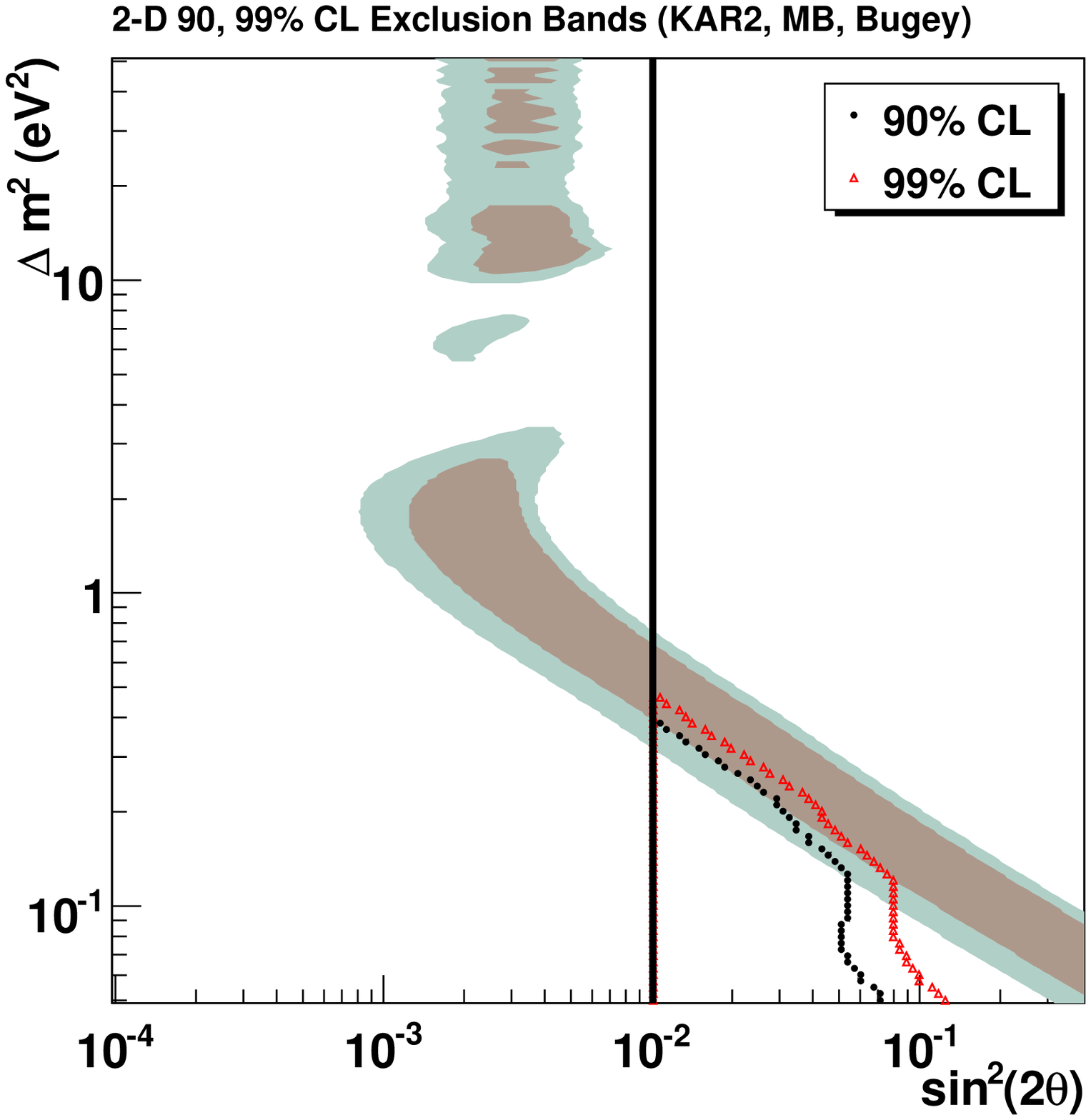}}
\caption{Top: Summed 2D $\Delta\chi^2$ compatibility grid from KARMEN2, MiniBooNE, and Bugey.  The star indicates the point of maximal compatibility (27.37\%).  Bottom: Exclusion bands (90\%, 99\%) found for the 2D KARMEN2, MiniBooNE, and Bugey joint analysis.  Values to the right of the lines are excluded at the 90, 99\% CL.  Triangle points form the 99\% CL band, circle points form the 90\% CL band.  The vertical straight edge on the left indicates the lower $\sin^2 2\theta$ bound of 0.01.}
\label{fig:mbkarbugregions_2d}
\end{figure}

If we ignore the positive LSND result, but now include Bugey, 
it is 27.37\% probable that we would have found all three null results in 
a world with two-neutrino ($\dmsq >$ 0.0488 $\evsq$) oscillations.  Please note that the point of maximal 
compatibility is limited by the boundary of the analyzed region.  
Figure~\ref{fig:mbkarbugregions_2d} (top) shows the 
2D compatibility of all three null results.  Figure~\ref{fig:mbkarbugregions_2d} (bottom) presents the 
remaining allowed regions.  The straight line on the left hand side is an artifact of the requirement that the analysis be performed over 
regions of phase space common to all experiments.

\begin{figure}[h]
\scalebox{0.35}{\includegraphics[angle=0]{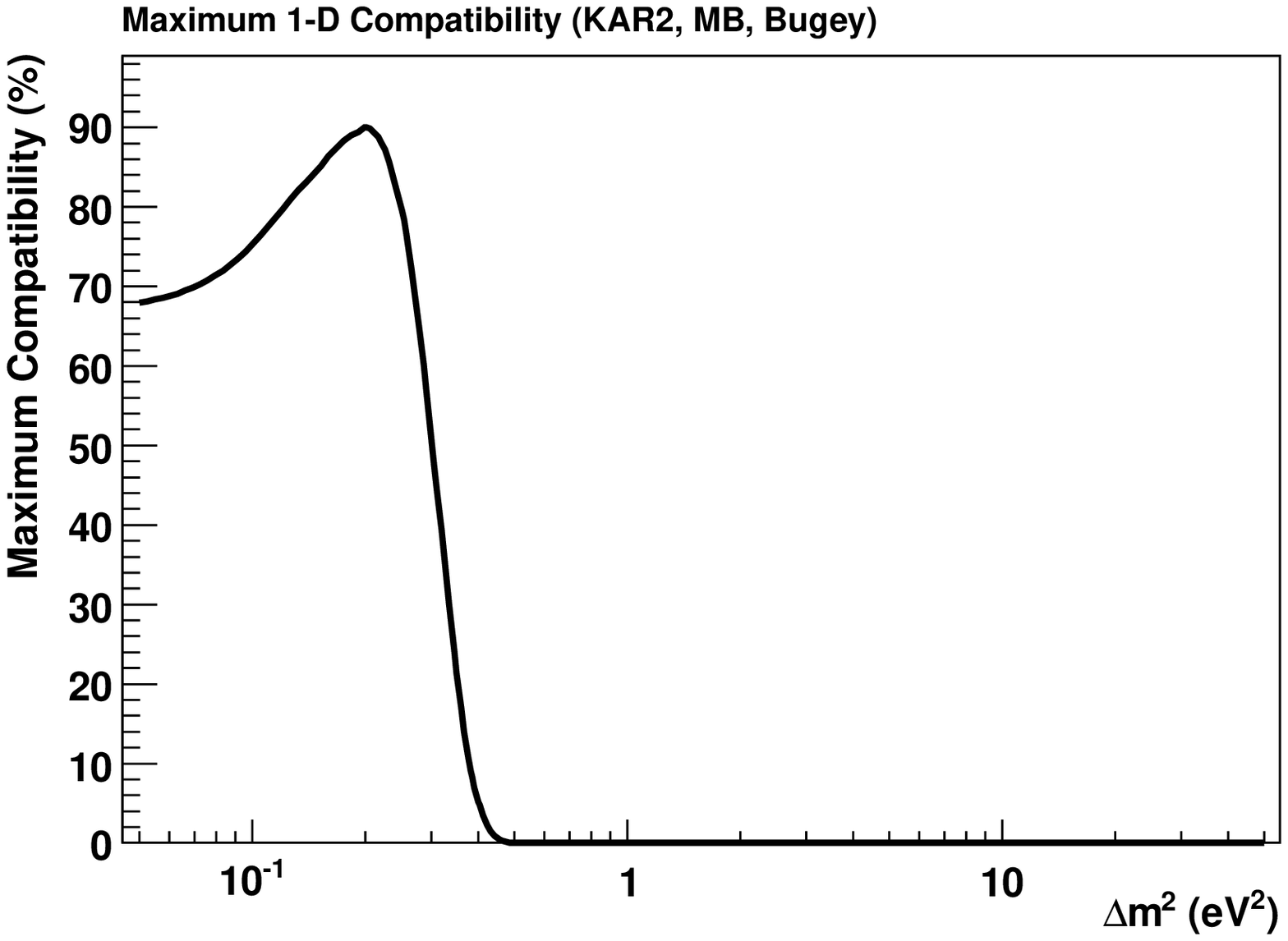}}
\hfill
\scalebox{0.35}{\includegraphics[angle=0]{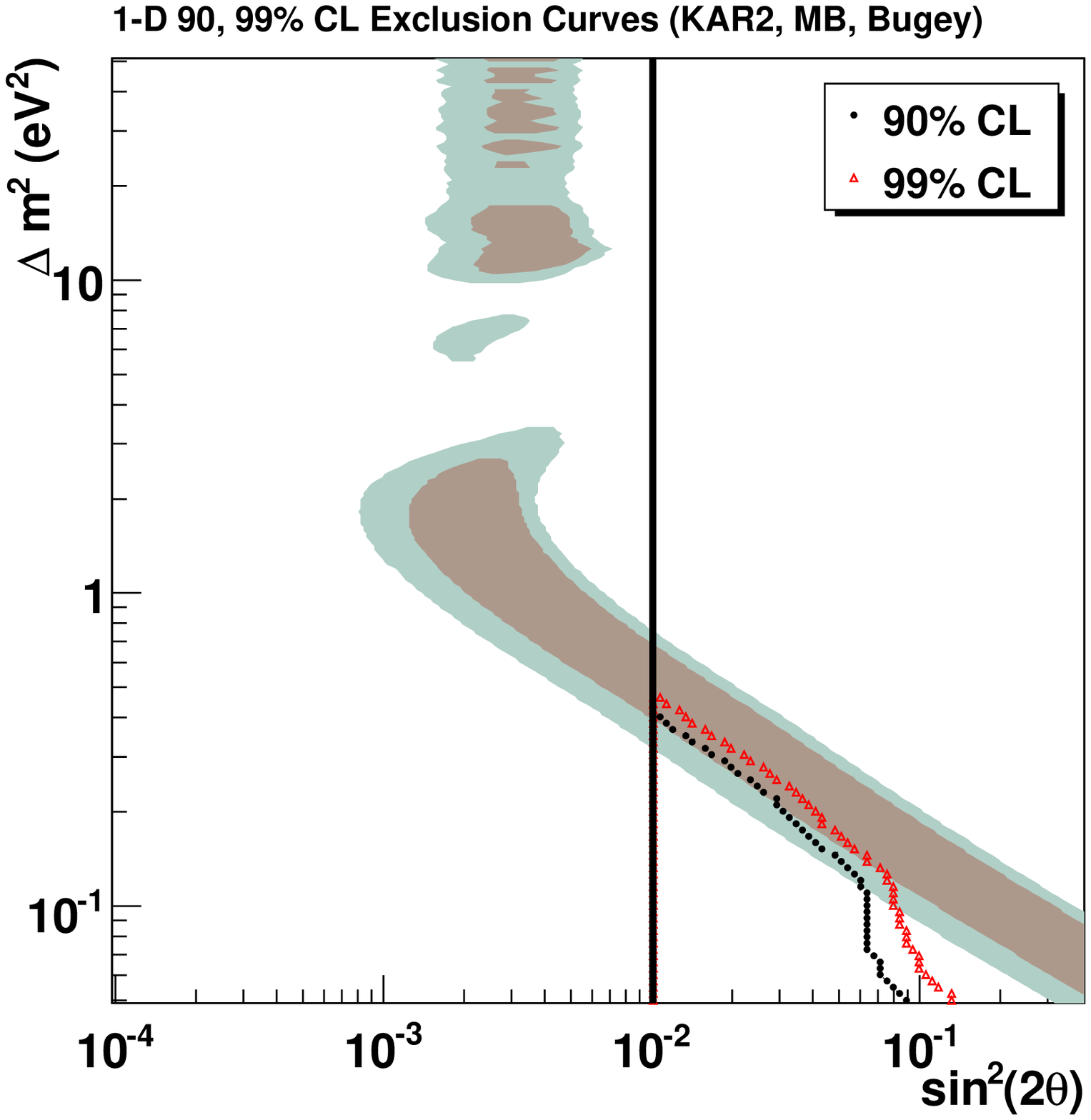}}
\caption{Top: Maximum compatibility as a function of $\Delta m^2$ for the 1D KARMEN2, MiniBooNE, and Bugey analysis.  Bottom: Exclusion bands (90\%, 99\%) found for the 1D KARMEN2, MiniBooNE, Bugey joint analysis.  Values to the right of the lines are excluded at the 90, 99\% CL.  Triangle points form the 99\% CL band, circle points form the 90\% CL band.  The vertical straight edge on the left indicates the lower $\sin^2 2\theta$ bound of 0.01.}
\label{fig:mbkarbugregions_1d}
\end{figure}

The 1D analysis of KARMEN2, MiniBooNE, and Bugey (top, Figure~\ref{fig:mbkarbugregions_1d})
 produces a higher degree of compatibility than the 2D analysis shown in Table~\ref{table:maxcompat_2d}, 
but agrees with the remaining allowed regions (bottom, Figure~\ref{fig:mbkarbugregions_1d}).

\subsection{LSND, MiniBooNE}
\begin{figure}[h]
\scalebox{0.35}{\includegraphics[angle=0]{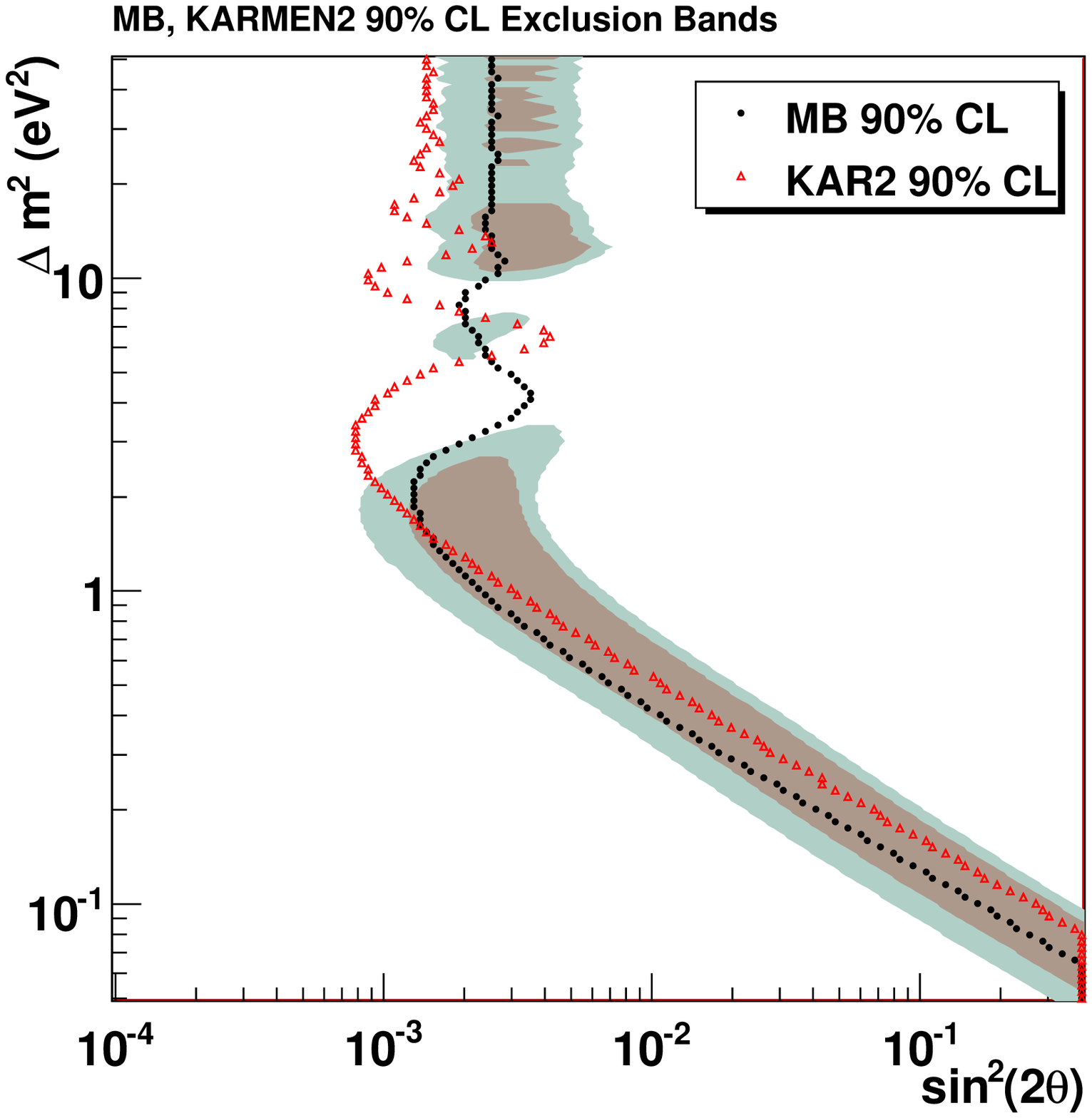}}
\caption{LSND 90, 99\% CL allowed regions, overlayed with the KARMEN2 (triangles) and MiniBooNE (circles) 2D 90\% CL 
exclusion bands.  KARMEN2 is more powerful at excluding high $\dmsq$ values while MiniBooNE is more powerful in
 the low $\dmsq$ region.}
\label{fig:mblsndkaroverlay}
\end{figure}

Table~\ref{table:maxcompat_2d} presents results from the combination of LSND and MiniBooNE, not including the 
KARMEN2 result.  The compatibility from the 2D analysis is actually lower 
than that found from the combination of LSND, KARMEN2, and MiniBooNE.  KARMEN2 and MiniBooNE are complementary 
results; KARMEN2 has the most power in high $\dmsq$ regions, while MiniBooNE is most sensitive to the 
lower $\dmsq$ areas.  The maximum compatibility of the (LSND, MiniBooNE), and (LSND, KARMEN2, MiniBooNE) analyses 
is found in the low $\dmsq$ region where MiniBooNE has the most power.  
The inclusion of KARMEN2 data  adds two degrees of freedom, 
but very little resolving power in this area of phase space.  Figure~\ref{fig:mblsndkaroverlay} illustrates this 
effect by overlaying the 2D LSND 90 and 99\% allowed regions with the MiniBooNE and KARMEN2 2D 
90\% exclusion curves.

MiniBooNE previously reported a 2\% compatibility for the combination of LSND and MiniBooNE, found using the 1D 
raster scan method~\cite{miniboone}.  The prior result was calculated over a restricted $\dmsq$ range (0.2 to 0.7 $\evsq$).
The current analysis, when restricted to the same $\dmsq$ range, agrees with the previously published result.

\subsection{LSND, KARMEN2}

The maximum compatibility of LSND and KARMEN2, found using the 2D method, is 32.21\%.  This differs 
from a previously reported compatibility of 64\%~\cite{joint}.  There are two differences between the current analysis 
and the study by Church {\em et~al.}: the input LSND data set (this analysis utilizes the 
LSND decay-in-flight and decay-at-rest results, while the previous analysis 
only used the LSND decay-at-rest data), and the method used to define and 
calculate the compatibility.  Both analyses find a high compatibility between LSND and KARMEN2.

\section{Conclusions}

We present results on the compatibility of different combinations of four experiments which have 
searched for neutrino oscillations at the high $\dmsq$ scale ($>$ 0.0488 $\evsq$).  
The LSND experiment has observed a significant excess of events; the other three experiments report 
null results and set limits on the oscillation parameter space.  The compatibility has been calculated using 
both a 2D and a 1D scan technique with the method of Reference~\cite{msch}.  The remaining allowed regions have been 
found for combinations resulting in greater than 10\% compatibility.  
Results from the 2D scan indicate that LSND, KARMEN2, and MiniBooNE are 
25.36\% compatible with having come from two-neutrino oscillations.  However, the best fit point for 
this analysis is found in a region excluded by Bugey.  (This point is also excluded by other reactor experiments such 
as Goesgen~\cite{goesgen}, and Krasnoyarsk~\cite{kras}.)  
The 2D scan from all four experiments including Bugey, 
in a limited $\sinsqtheta$ region common to all experiments, finds they are only 3.94\% compatible 
with two-neutrino oscillations.  This analysis does not take into consideration the absolute goodness of fit of 
each individual experiment at its own best fit point, or any additional non-standard 
model effects such as CP violation or sterile neutrinos. \\

\begin{acknowledgments}
We acknowledge the support of Fermilab, the Department of Energy,
and the National Science Foundation. We thank M. Acero, C. Giunti, and M. Laveder for 
providing us with the Bugey data.  We thank K. Eitel for providing us with the KARMEN2 data.
\end{acknowledgments}

\bibliography{prl}

\end{document}